\documentclass[prb,superscriptaddress,amsmath,amssymb,twocolumn]{revtex4-1}
\usepackage{amsmath,amssymb}
\usepackage{graphicx}
\usepackage{bm}
\usepackage{mathtools}
\usepackage[format=plain,labelfont=bf,font=small,justification=raggedright]{caption}
\usepackage{xcolor}

\usepackage{atbegshi}
  \AtBeginShipoutFirst{}
\usepackage[colorlinks=true,linkcolor=blue,citecolor=blue,anchorcolor=blue,urlcolor=blue,pdfborderstyle={}]{hyperref}

\DeclareCaptionLabelSeparator{bar}{~$|$~}

\newcommand{\panel}[1]{\large{\textbf{#1}}}
\newcommand{\bb}[1]{\textbf{#1}}

\newcommand{\ri}{\mathrm{i}}  
\newcommand{\e}{\mathrm{e}}
\newcommand{\vecr}{{\bm{r}}}
\renewcommand{\eqref}[1]{Eq.~(\ref{#1})}

\newcommand{\pref}[1]{(\ref{#1})}
\newcommand{\figref}[1]{Fig.~\ref{#1}}

\begin{document}
\title{Universal Critical Behavior at a Phase Transition to Quantum Turbulence}

\author{Masahiro Takahashi}
\affiliation{Department of Physics, Gakushuin University, Tokyo 171-8588, Japan}
\author{Michikazu Kobayashi}
\affiliation{Department of Physics, Kyoto University, Oiwake-cho, Kitashirakawa, Sakyo-ku, Kyoto 606-8502, Japan} 
\author{Kazumasa A. Takeuchi}
\email{kat@kaztake.org}
\affiliation{Department of Physics, Tokyo Institute of Technology, 2-12-1-H85 Ookayama, Meguro-ku, Tokyo 152-8551, Japan}
\date{\today}


\begin{abstract}
Turbulence is one of the most prototypical phenomena of systems driven out of equilibrium.
While turbulence has been studied mainly with classical fluids like water,
 considerable attention is now drawn to quantum turbulence (QT),
 observed in quantum fluids
 such as superfluid helium and Bose-Einstein condensates.
A distinct feature of QT is that
 it consists of quantum vortices, by which turbulent circulation is quantized.
Yet, under strong forcing,
 characteristic properties of developed classical turbulence
 such as Kolmogorov's law
 have also been identified in QT.
Here, we study the opposite limit of weak forcing,
 i.e., the onset of QT, numerically,
 and find
 another set of universal scaling laws
 known for classical non-equilibrium systems.
Specifically, we show that the transition belongs
 to the directed percolation universality class,
 known to arise generically in transitions into an absorbing state,
 including transitions to classical shear-flow turbulence
 after very recent studies.
We argue that quantum vortices play an important role in our finding.
\end{abstract}
\maketitle


Studies on quantum turbulence (QT), which had long focused on thermal counterflow
 in superfluid helium \cite{Tough-Inbook1982},
 entered a new phase in 1990s.
Firstly, atomic BECs were experimentally realized
 \cite{Pethick.Smith-Book2008}.
With Bose-Einstein condensates (BECs), one can not only visualize the system directly
 but control the trap potential and interactions \cite{Pethick.Smith-Book2008},
 leading to recent realizations of QT in BECs
 \cite{Henn.etal-PRL2009,Neely.etal-PRL2013,Kwon.etal-PRA2014,Tsatsos.etal-a2015}.
Secondly, growing interests have been aroused
 in analogy to classical turbulence
 \cite{Vinen.Niemela-JLTP2002,Vinen.Donnelly-PT2007,Tsubota.etal-PR2013}.
In particular, Kolmogorov's $-5/3$ law
 for developed turbulence \cite{Frisch-Book1995},
 $E(k)\sim{}k^{-5/3}$, with wave number $k$ and energy spectrum $E(k)$,
 has been also found in QT, both experimentally
 \cite{Maurer.Tabeling-EL1998,Salort.etal-PF2010}
 and numerically
 \cite{Nore.etal-PF1997,Kobayashi.Tsubota-PRL2005,Sasa.etal-PRB2011,Kobayashi.Tsubota-JLTP2006},
 despite the quite different nature of the turbulent eddies.

In contrast to such progress on developed QT,
 less is understood on its onset,
 characterized by the generation of turbulent quantum vortices.
Transitions to QT
 have been observed in thermal counterflow \cite{Tough-Inbook1982},
 superfluid driven by oscillatory obstacles \cite{Hanninen.etal-PRB2007}, and trap-driven BECs \cite{Henn.etal-PRL2009,Neely.etal-PRL2013,Kwon.etal-PRA2014,Tsatsos.etal-a2015},
 but there seems to be no common approach to deal with statistical properties
 of these transitions.
On the other hand, from broader perspectives,
 they can be regarded as phase transitions
 in systems driven out of equilibrium.
Among various kinds of non-equilibrium phase transitions,
 one of the most important is transitions
 into an absorbing state \cite{Hinrichsen-AP2000},
 for which a number of universality classes have been established.
Here, the absorbing state refers to a global state that
 the system can enter but can never leave.
One can actually expect a similar state to exist in QT transitions,
 because once all quantum vortices disappear,
 their spontaneous nucleation is basically hindered by its high energy barrier.
Physics of absorbing-state transitions may therefore provide a useful viewpoint
 to analyze and elucidate QT transitions in general.

This direction is pursued in this Letter.
We numerically study the Gross-Pitaevskii (GP) equation, known to provide
 a quantitative description of BECs \cite{Pethick.Smith-Book2008}.
With a phenomenological dissipation term, the GP equation reads
\begin{equation}
(\ri - \gamma) \frac{\partial \psi}{\partial t} = - \bm{\nabla}^2 \psi + (V_{\vecr,t} - \mu) \psi + g | \psi |^2 \psi,  \label{eq:GP}
\end{equation}
 with macroscopic wave function $\psi(\vecr,t)$,
 external potential $V_{\vecr,t}$, chemical potential $\mu$,
 coupling constant $g$, dissipation coefficient $\gamma$, and
 $\hbar=2m=1$ where $m$ is the atomic mass.
We use $\gamma = \mu = g = 1$
 and the periodic boundary condition.

\begin{figure*}[t]
\hspace{-.05\hsize}
 \begin{minipage}{.35\hsize}
  \centering  \panel{a}\\
  \includegraphics[width=.8\hsize]{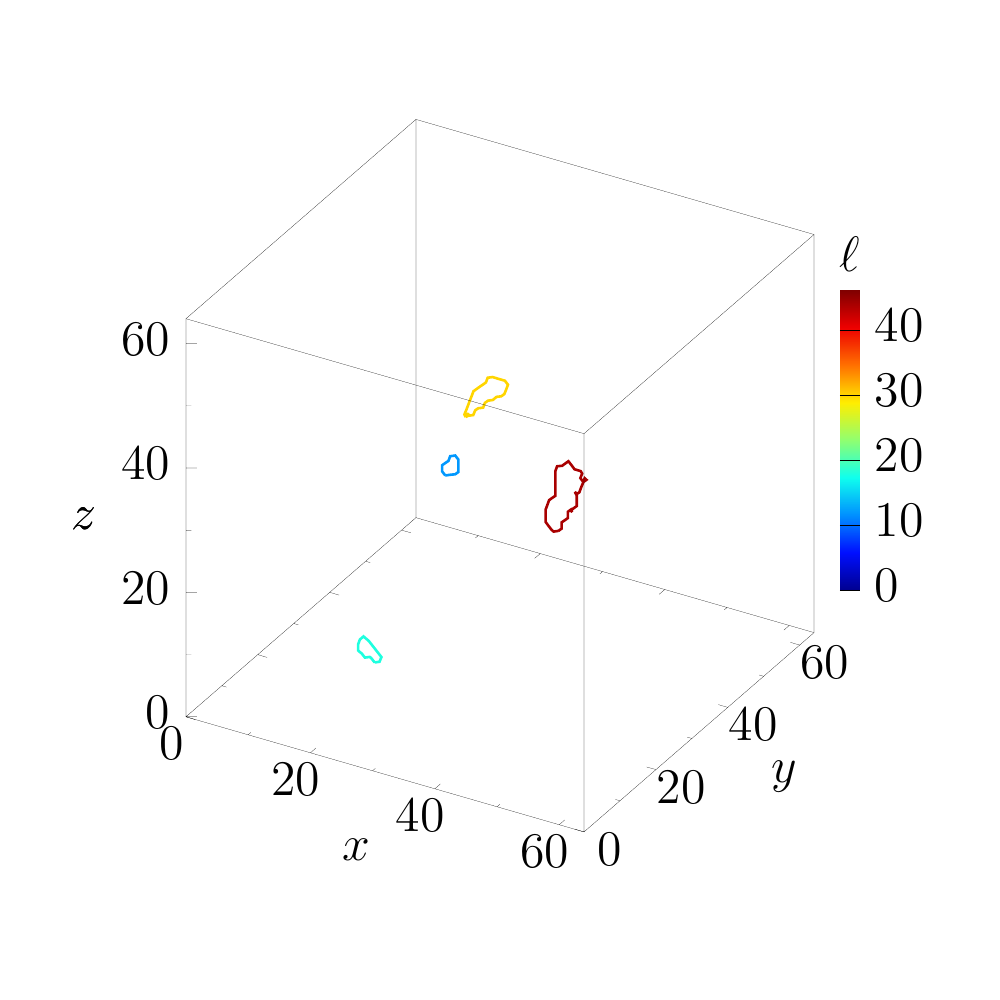}
 \end{minipage}
\hspace{-.1\hsize}
 \begin{minipage}{.35\hsize}
  \centering  \panel{b}\\
  \includegraphics[width=.8\hsize]{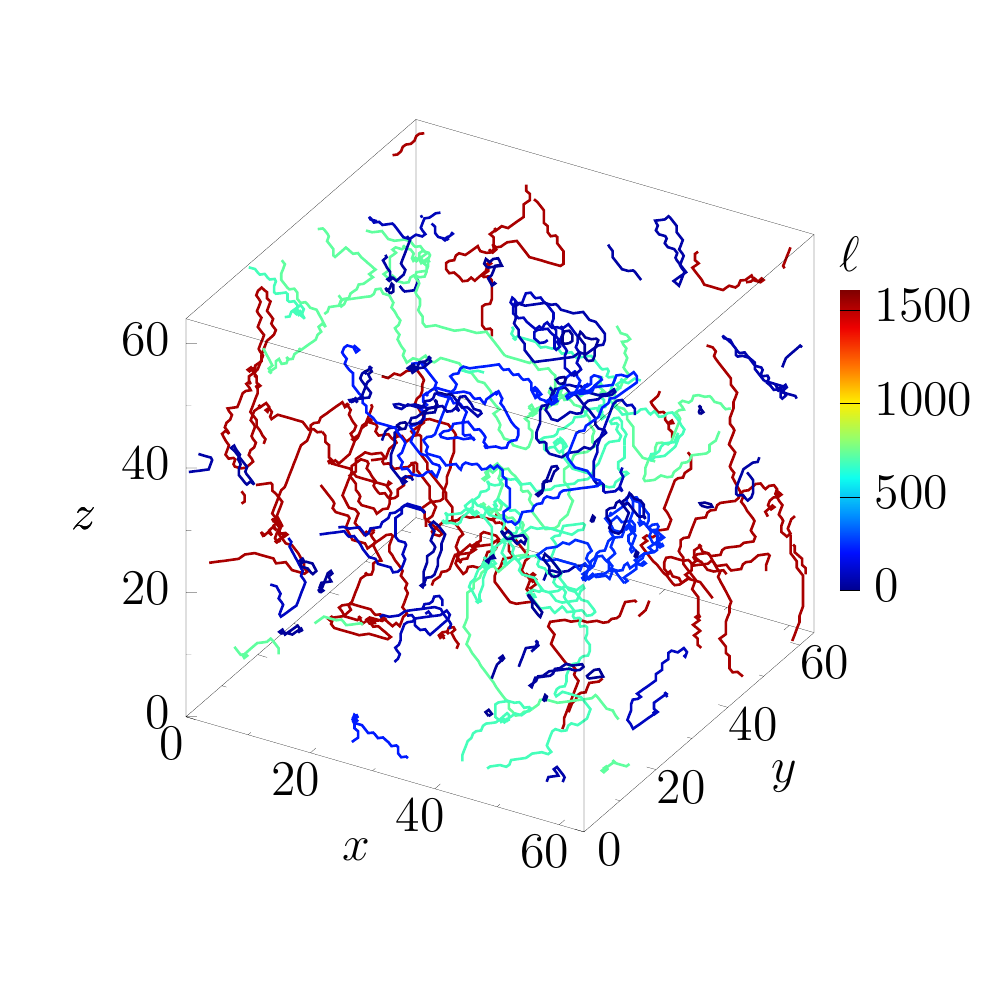}
 \end{minipage}
 \begin{minipage}{.35\hsize}
  \centering  \panel{c}\\
  \includegraphics[width=\hsize]{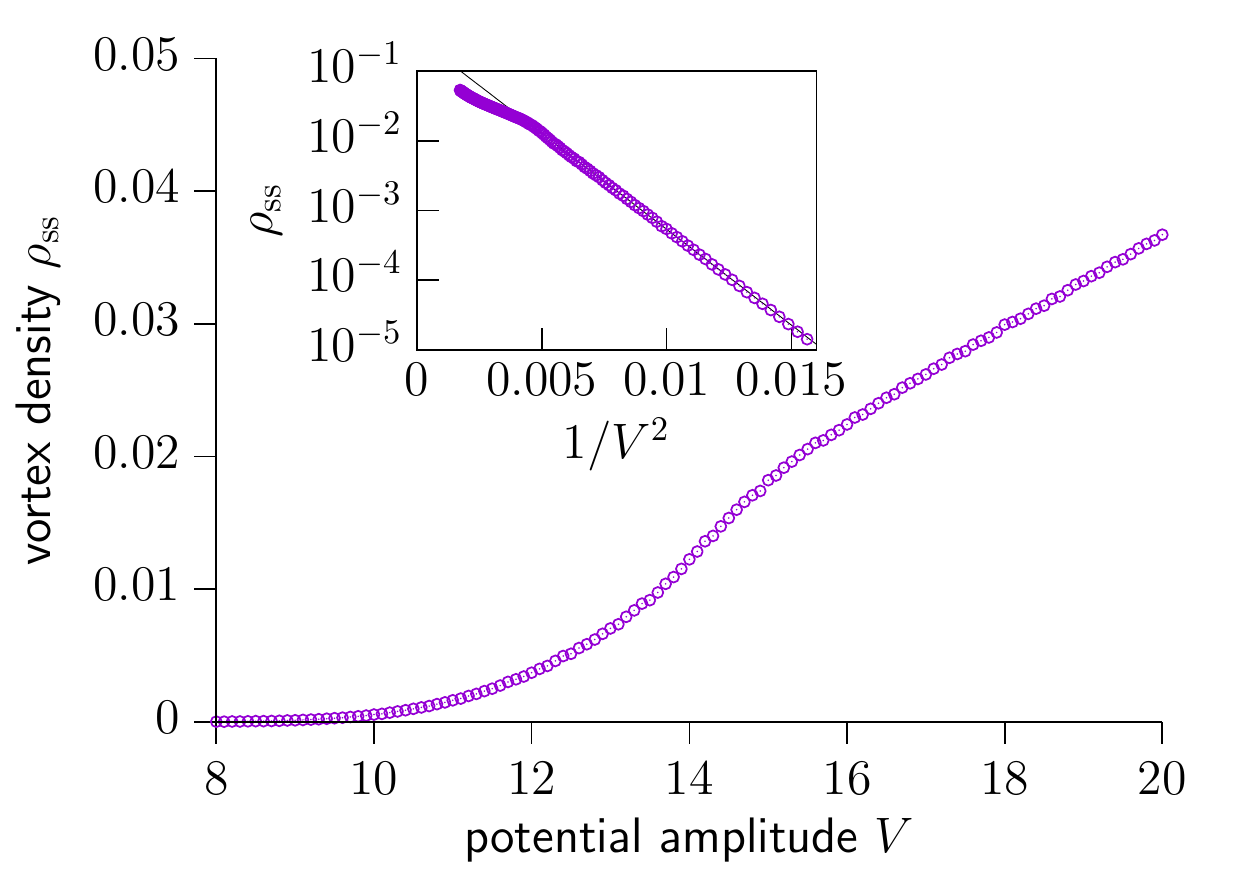}
 \end{minipage}
 \caption{
\bb{Overview of the QT transition in the GP equation.}
\bb{a},\bb{b}, Snapshots of the steady state at $V = 10 (<V_\mathrm{c})$
 (\bb{a}) and $V = 15 (>V_\mathrm{c})$ (\bb{b}).
The color indicates the length $\ell$ of each vortex.
See also Supplementary Video 1.
\bb{c}, Steady-state vortex density $\rho_\mathrm{ss}(V)$
 against the potential amplitude $V$.
The inset shows the same data against $1/V^2$.
The black line indicates a fit to $\e^{-c_\rho/V^2}$.
}
 \label{fig1}
\end{figure*}%

\begin{figure*}[t]
 \begin{minipage}{.45\hsize}
  \centering  \panel{a}\\
  \includegraphics[width=\hsize]{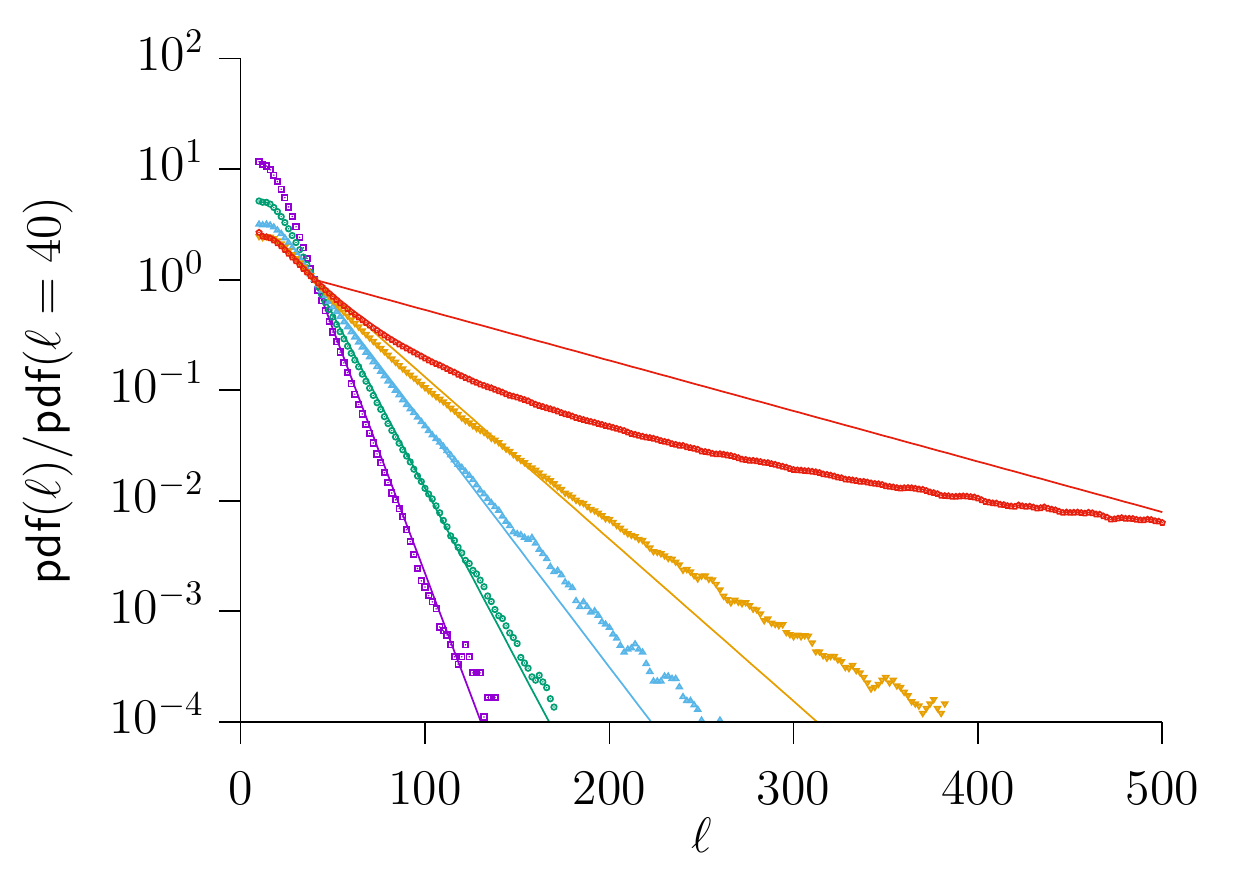}
 \end{minipage}
 \begin{minipage}{.45\hsize}
  \centering  \panel{b}\\
  \includegraphics[width=\hsize]{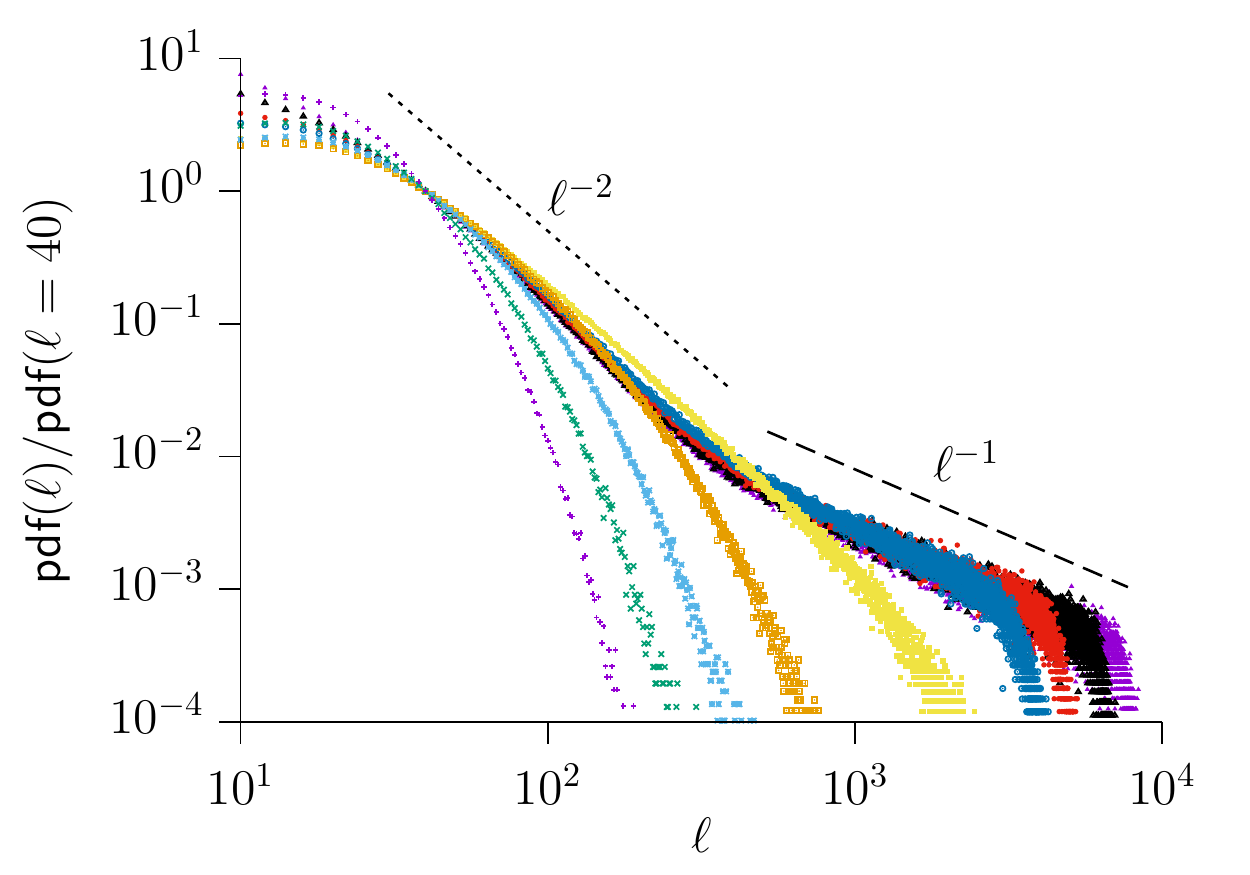}
 \end{minipage}
 \begin{minipage}{.45\hsize}
  \centering  \panel{c}\\
  \includegraphics[width=\hsize]{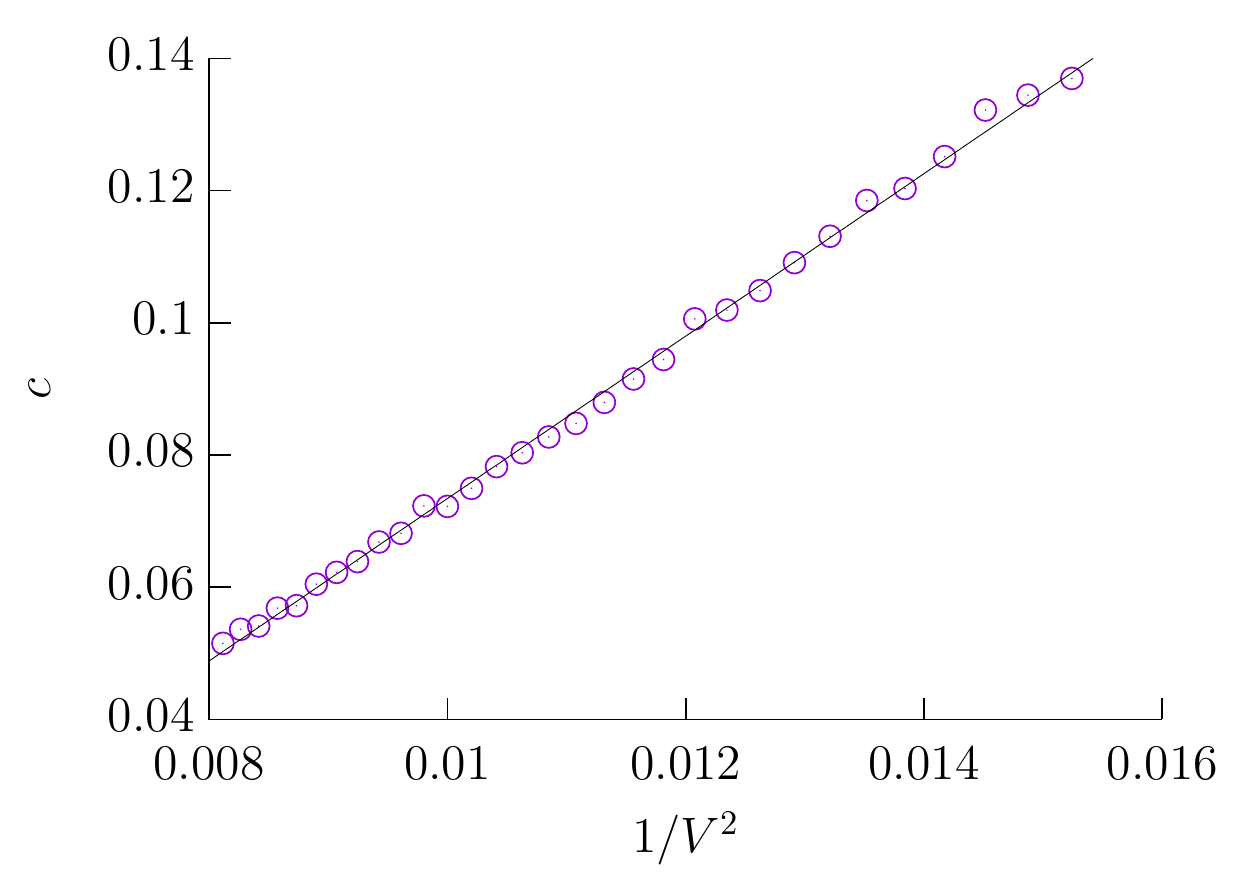}
 \end{minipage}
 \begin{minipage}{.45\hsize}
  \centering  \panel{d}\\
  \includegraphics[width=\hsize]{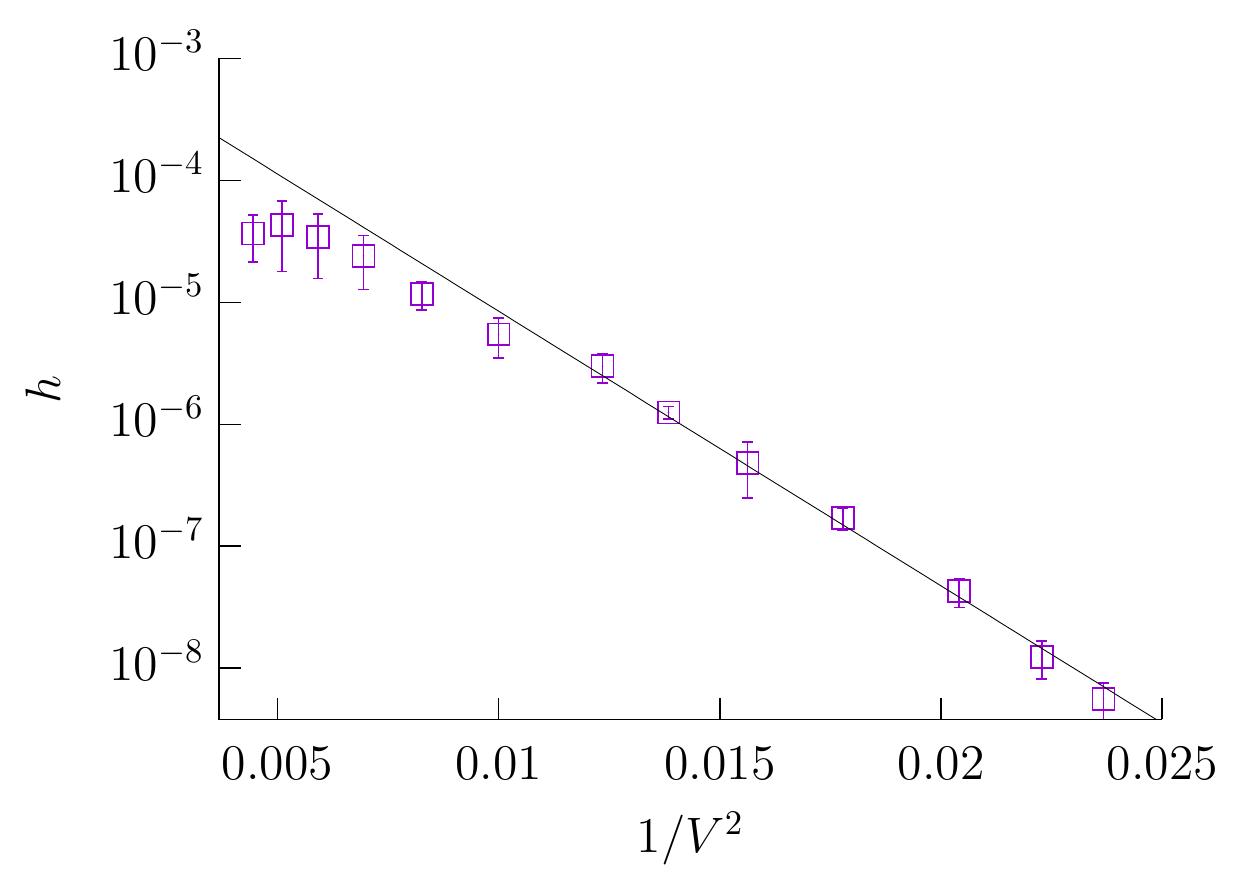}
 \end{minipage}\\
 \caption{
\bb{Thermal and nonthermal effects.}
\bb{a,b}, 
Distribution (pdf) of vortex length $\ell$
 for $V=9,10,11,12,14$ (\bb{a}, semi-log plots)
 and for $V=10,11,12,13,14,15,16,18,20$ (\bb{b}, log-log plots)
 from left to right.
The lines in the panel \bb{a}
 indicate the Boltzmann distribution $\e^{-c\ell}$
 with the value of $c$ determined
 from the regression shown in the panel \bb{c}.
The dotted and dashed lines in the panel \bb{b}
 are guides for the eyes indicating exponent $-2$ and $-1$, respectively.
\bb{c}, $c$ against $1/V^2$. 
The data points are obtained by fitting $\e^{-c\ell}$
 to the exponential region of pdf$(\ell)$.
The line shows the result of the linear regression.
\bb{d}, Spontaneous nucleation rate of vortices per unit volume and time,
 $h(V)$, shown against $1/V^2$.
The black line indicates a fit to $\e^{-c_h/V^2}$.
}
 \label{fig2}
\end{figure*}%

To study QT, we drive the system by random fluctuations
 of the external potential $V_{\vecr,t}$, prepared in space and time
 with correlation length $\ell_\mathrm{v} = 8$,
 correlation time $t_\mathrm{v} = 0.1$, and amplitude $V$
 (see Methods for details).
For low $V$, quantum vortices are occasionally generated,
 but they disappear soon as a result of dissipation
 (\figref{fig1}a and Supplementary Video 1 left).
In contrast, when $V$ is sufficiently elevated,
 there remain some fluctuating vortices
 (\figref{fig1}b and Supplementary Video 1 right),
 maintaining a constant density on average.
This observation can be quantified by measuring
 the steady-state vortex density $\rho_\mathrm{ss}(V)$,
 which is found to grow significantly around $10\lesssim{}V\lesssim{}15$
 (\figref{fig1}c).
While this implies a transition,
 $\rho_\mathrm{ss}(V)$ is found not to vanish even at very low $V$,
instead decaying as $\e^{-c_\rho/V^2}$ with constant $c_\rho$ (inset).

The presence of vortices at low $V$ can be understood as thermal effect
 due to the random potential $V_{\vecr,t}$,
 with effective temperature $T_\mathrm{eff}$
 proportional to $V^2$ (i.e., variance).
Figures~\ref{fig2}a and \ref{fig2}b show
 the steady-state distribution of vortex lengths $\ell$ at different $V$.
For $V \lesssim 10$, we find the Boltzmann distribution $\e^{-c\ell}$,
 with a $V$-dependent coefficient
 $c(V) \sim 1/T_\mathrm{eff} \sim 1/V^2$
 (\figref{fig2}c) expected for thermal effect.
By contrast, from $V \approx 11$
 the distribution deviates from Boltzmann (\figref{fig2}a),
 first developing a power-law tail with exponent $-2$
 and eventually giving rise to another tail with exponent $-1$
 (\figref{fig2}b).
This implies that nonthermal effect plays a crucial role for $V \gtrsim 11$.
We also measure the local nucleation rate of vortices, $h(V)$,
 and confirm the expected Arrhenius law $h \sim \e^{-c'/V^2}$
 for $V \lesssim 10$ (\figref{fig2}d).
For higher $V$, $h(V)$ appears to be somewhat lower
 than the Arrhenius law would predict, but no singularity is identified.
Thermal nucleation has therefore only secondary effect in our QT transition.

\begin{figure*}[t]
 \begin{minipage}{.45\hsize}
  \centering  \panel{a}\\
  \includegraphics[width=\hsize]{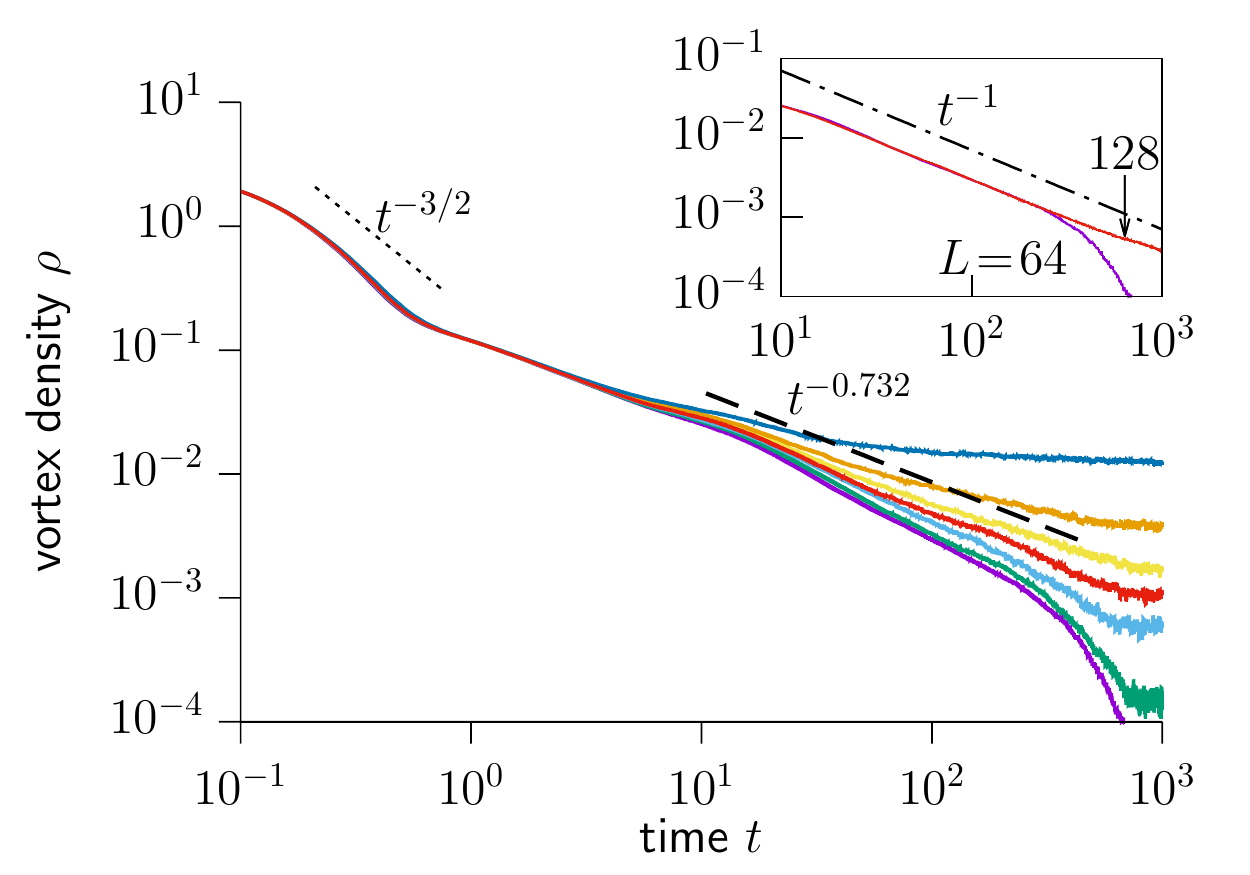}
 \end{minipage}
 \begin{minipage}{.45\hsize}
  \centering  \panel{b}\\
  \includegraphics[width=\hsize]{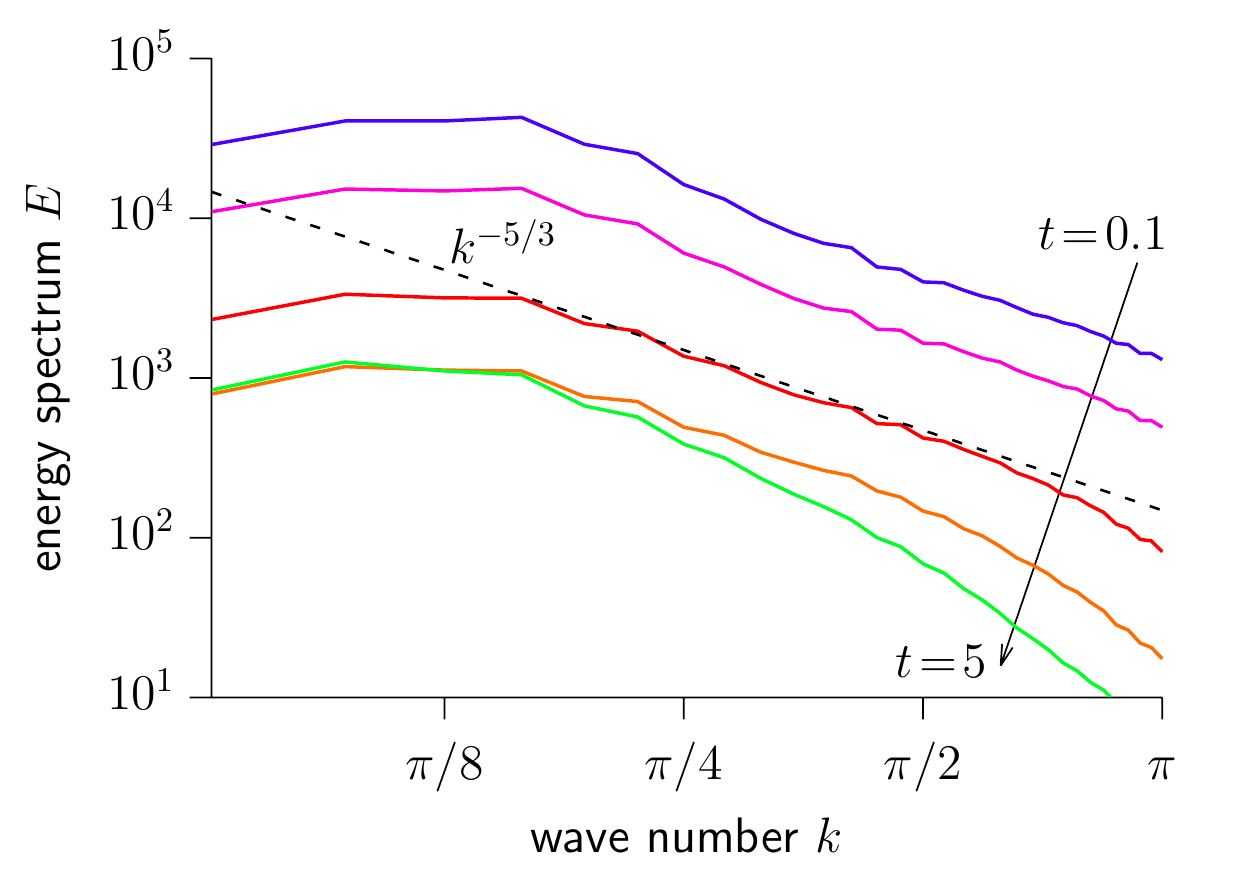}
 \end{minipage}
 \caption{
\bb{Results of quench simulations.}
\bb{a}, 
Vortex density $\rho(t)$
 for $V=8,9,10,10.5,11,12,14$ (from bottom to top).
The critical case $V=V_\mathrm{c}=10.5$ is drawn in red.
The dashed and dotted lines are guides for the eyes
 indicating the DP critical decay $t^{-\alpha}$
 with $\alpha^\mathrm{DP}=0.732$ and the Kolmogorov decay
 $t^{-3/2}$, respectively.
The inset shows the decay at $V=8 < V_\mathrm{c}$
 for $L=64$ (purple) and 128 (red), compared with $t^{-1}$
 for the standard coarsening process of quantum vortices.
\bb{b}, 
Incompressible part of the kinetic energy spectrum $E(k)$ for $V=V_\mathrm{c}$
 at $t=0.1, 0.2, 0.5, 1, 5$ as indicated by the arrow.
The dashed line indicates Kolmogorov's law $E(k) \sim k^{-5/3}$.
}
 \label{fig3}
\end{figure*}%

To better characterize the transition,
 we study turbulence decay after a sudden drop of the potential amplitude,
 from $V=140$ corresponding to the regime of developed QT,
 to a target value $V$ chosen near the critical point
 (see Supplementary Video 2 for vortex dynamics at $V=10.5$).
Figure~\ref{fig3}a shows the vortex density $\rho(t)$
 -- our order parameter -- obtained by such quench simulations
 for different $V$ at system size $L=64$.
While the curves $\rho(t)$ mostly overlap up to $t \approx 10$,
 eventually they split, accelerating (low $V$) or decelerating (high $V$)
 before reaching the steady-state value $\rho_\mathrm{ss}(V)$.
This allows us to determine the QT critical point unambiguously,
 at $V_\mathrm{c}=10.5(5)$,
 where the numbers in parentheses
 indicate the range of errors in the last digit.
Moreover, at $V=V_\mathrm{c}$ (red curve in \figref{fig3}a),
 we find clear power-law decay
\begin{equation}
 \rho(t) \sim t^{-\alpha} \quad (V=V_\mathrm{c}).  \label{eq:Quench}
\end{equation}
 with $\alpha=0.74(4)$.
Remarkably, this value of $\alpha$ shows
 good agreement with that of the (3+1)-dimensional directed percolation (DP)
 universality class \cite{Hinrichsen-AP2000},
 $\alpha^\mathrm{DP}=0.732(4)$ \cite{Jensen-PRA1992}.
The DP class is known to describe
 the simplest type of absorbing-state transitions
 without extra symmetry or conservation \cite{Hinrichsen-AP2000},
 with firm theoretical basis \cite{Hinrichsen-AP2000}
 and a few experimental examples \cite{Takeuchi.etal-PRL2007,Takeuchi.etal-PRE2009,Sano.Tamai-NP2016,Lemoult.etal-NP2016}.
In our system, the nucleation rate of vortices (\figref{fig2}d) is low enough
 that such an absorbing state is approximately realized.

In fact, interests of our quench results
 are not restricted to the DP critical behavior,
 which asymptotically arises at $V=V_\mathrm{c}$.
For example, at earlier times, we notice that $\rho(t)$ decays
 faster than \eqref{eq:Quench},
 roughly following $\rho(t) \sim t^{-3/2}$ (dotted line in \figref{fig3}a)
 albeit the very narrow time window.
In fact, this exponent indicates characteristic decay from developed QT
 governed by Kolmogorov scaling laws
 \cite{Smith.etal-PRL1993,Kobayashi.Tsubota-JLTP2006,Stalp.etal-PRL1999,Bradley.etal-PRL2006}.
To test it, we plot in \figref{fig3}b
 the incompressible part of the kinetic energy spectrum
 $E(k)$ at different times,
 and indeed find the Kolmogorov exponent $-5/3$ until $t \lesssim 1$,
 while the $t^{-3/2}$ decay is present.
Based on these observations, we conclude that the relaxation
 at early times is indeed governed by the Kolmogorov scaling scenario,
 which is then replaced by the DP critical behavior at late times
 for $V=V_\mathrm{c}$.

Another remark concerns the case of $V<V_\mathrm{c}$.
According to the standard scaling hypothesis
 for absorbing-state transitions \cite{Hinrichsen-AP2000},
 below $V_\mathrm{c}$, $\rho(t)$ asymptotically decays exponentially.
Although this seems to be consistent with our observation in \figref{fig3}a,
 obtained at $L=64$, for larger system sizes,
 we find instead $\rho(t)\sim{}t^{-1}$ within the observation time (inset).
This power-law relaxation is actually a hallmark
 of the coarsening process expected for quantum vortices,
 observed previously at relaxation to thermal equilibrium
 \cite{Kobayashi.Tsubota-JLTP2006}.
The same behavior arises here, presumably for long but finite length of time,
 because of the finite system size $L$.
We then expect it to be replaced by the usual exponential decay,
 following the theoretical framework of the DP class,
 until $\rho(t)$ reaches $\rho_\mathrm{ss}(V)$
 which remains positive due to thermally induced vortices.

\begin{figure*}[t]
 \begin{minipage}{.325\hsize}
  \centering  \panel{a}\\
  \includegraphics[width=\hsize]{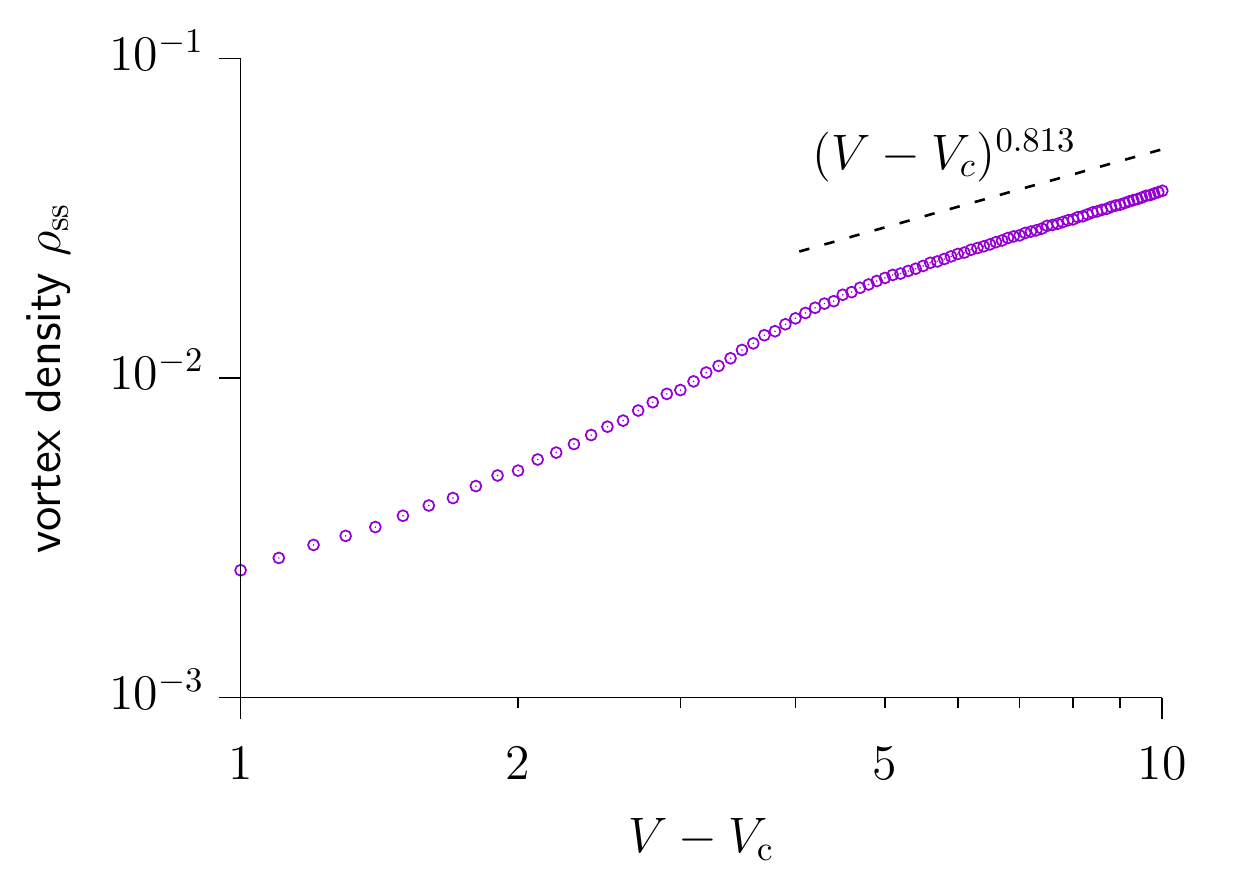}
 \end{minipage}
 \begin{minipage}{.325\hsize}
  \centering  \panel{b}\\
  \includegraphics[width=\hsize]{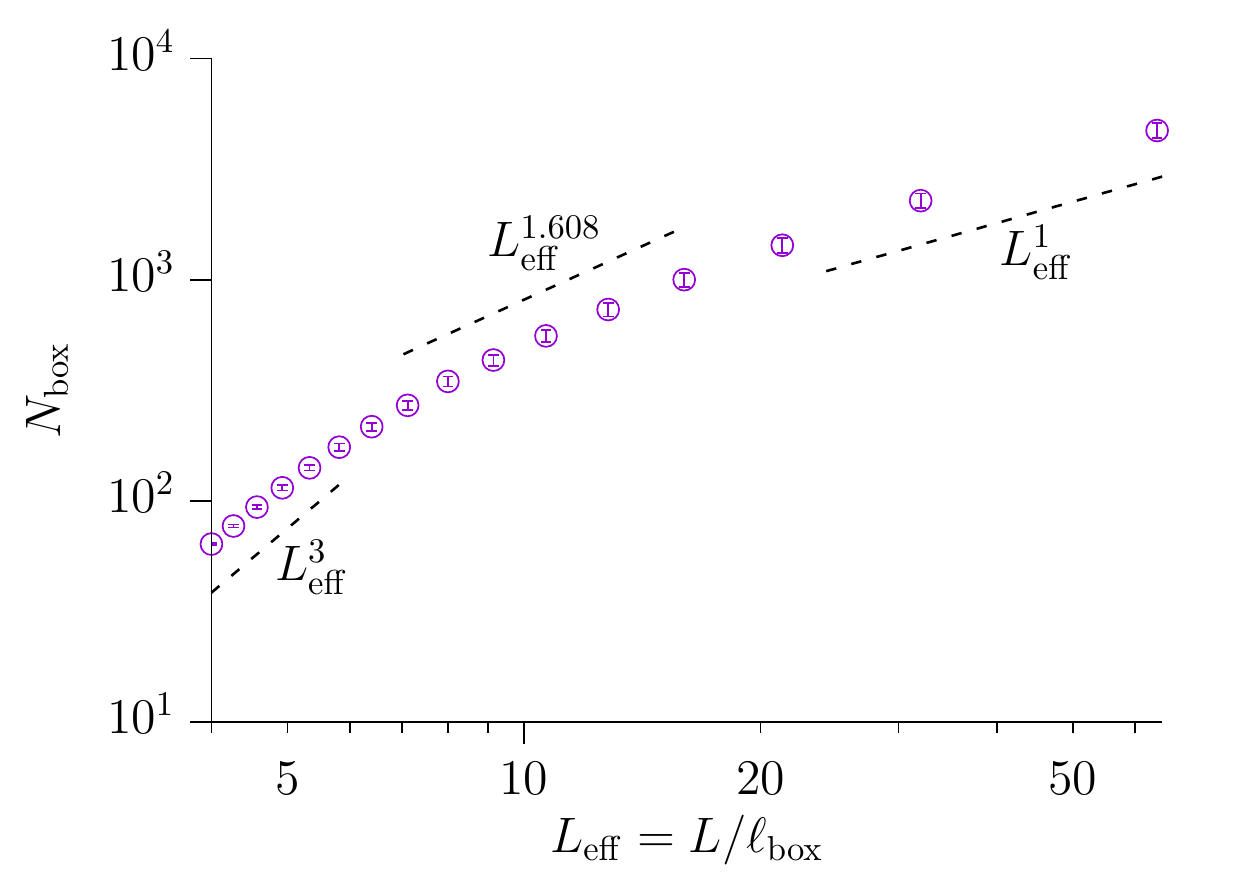}
 \end{minipage}
 \begin{minipage}{.325\hsize}
  \centering  \panel{c}\\
  \includegraphics[width=\hsize]{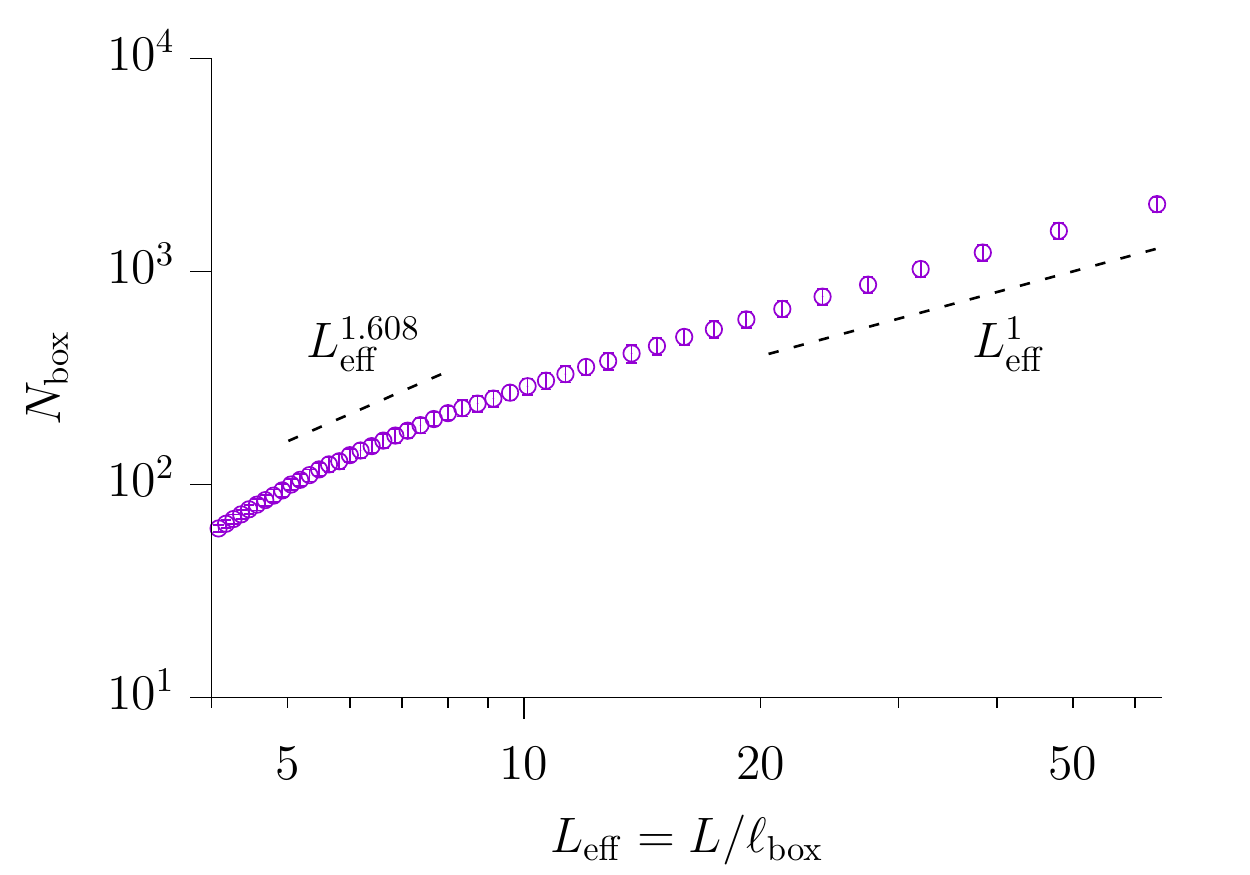}
 \end{minipage}\\
 \caption{
\bb{Order-parameter exponent $\beta$ and fractal dimension $d_\mathrm{f}$.}
\bb{a}, 
Steady-state vortex density $\rho_\mathrm{ss}(V)$.
The data in \figref{fig1}c are plotted in logarithmic scales
 against $V-V_\mathrm{c}$ with $V_\mathrm{c}=10.5$.
The dashed line indicated the DP scaling law \pref{eq:SteadyRho}
 with $\beta^\mathrm{DP}=0.813$.
\bb{b},\bb{c},
Number of boxes $N_\mathrm{box}$ of size $\ell_\mathrm{box}$
 that contain (any piece of) quantum vortices,
 shown as a function of the effective system size
 $L_\mathrm{eff}=L/\ell_\mathrm{box}$,
 in the steady state at $V=15$ (\bb{b})
 and in the quench simulations at $V=10.5$ and $t=50$ (\bb{c}).
The three different power laws are indicated by the dashed lines.
The one in the middle is characteristic of DP,
 $N \sim L_\mathrm{eff}^{d_f}$ with $d_\mathrm{f}^\mathrm{DP}=1.608$.
}
 \label{fig4}
\end{figure*}%

Theoretically, the DP class is characterized
 by three independent critical exponents \cite{Hinrichsen-AP2000}.
For example, the order-parameter exponent $\beta$ is defined by
\begin{equation}
 \rho_\mathrm{ss}(V) \sim (V-V_\mathrm{c})^\beta \quad (V > V_\mathrm{c})
  \label{eq:SteadyRho}
\end{equation}
 with the steady-state order parameter $\rho_\mathrm{ss}(V)$,
 which we already showed in \figref{fig1}c.
Plotting the same data against $V-V_\mathrm{c}$ in logarithmic scales,
 with $V_\mathrm{c}=10.5$ from the quench simulations,
 we indeed find agreement with $\beta^\mathrm{DP}=0.813(11)$
 \cite{Jensen-PRA1992} for the DP class (\figref{fig4}a).
Note that, somewhat peculiarly,
 the DP power law \pref{eq:SteadyRho} arises only for $V \gtrsim 15$,
 while $\rho_\mathrm{ss}(V)$ takes 
 values lower than the power law \pref{eq:SteadyRho} indicates
 for $V_\mathrm{c} \lesssim V \lesssim 15$.
Interestingly, $V \approx 15$ is also the threshold
 for the vortex-length distribution to develop the power-law tail
 with exponent $-1$ (\figref{fig2}b).

For the third independent exponent,
 we measure the fractal dimension of critical clusters,
 composed of quantum vortices.
Employing the standard box-counting method \cite{Barnsley-Book2012},
 we fill the whole space by boxes of size $\ell_\mathrm{box}$
 and count the number of boxes containing (any piece of) quantum vortices,
 $N_\mathrm{box}(\ell_\mathrm{box})$.
It is plotted
 in Figs.~\ref{fig4}b (steady state) and \ref{fig4}c (quench)
 as a function of the effective system size
 $L_\mathrm{eff}\equiv{}L/\ell_\mathrm{box}$.
The result shows (i) $N_\mathrm{box}\sim{}L_\mathrm{eff}$
 for large $L_\mathrm{eff}$ (small box sizes),
 reflecting the filamentous structure of vortices;
 (ii) $N_\mathrm{box}\sim{}L_\mathrm{eff}^3$ for small $L_\mathrm{eff}$
 (large box sizes), indicating the trivial spatial dimension $d=3$;
 (iii) in between, we find $N_\mathrm{box}\sim{}L_\mathrm{eff}^{d_\mathrm{f}}$
 with nontrivial exponent $d_\mathrm{f}\approx1.6$.
This can be regarded as the critical fractal dimension,
 which is indeed
 $d_\mathrm{f}^\mathrm{DP}=1.608(9)$
 for DP \cite{Takeuchi.etal-PRE2009,Jensen-PRA1992}
 in agreement with our estimate.

In summary, we have measured three critical exponents
 $\alpha, \beta, d_\mathrm{f}$ characterizing the QT transition
 in the GP equation,
 and found all of them in agreement with the DP universality class,
 known for non-equilibrium, absorbing-state transitions.
Since the DP class is characterized
 by three independent exponents \cite{Hinrichsen-AP2000},
 we can safely conclude that our QT transition belongs to the DP class.
Although there are numerous models known to be in the DP class
 \cite{Hinrichsen-AP2000},
 including a recent one for Rydberg gas \cite{Marcuzzi.etal-NJP2015},
 to our knowledge it is the first example
 where a DP-class transition is found in a generic situation of quantum fluids,
 without underlying lattice structure nor extra fine tuning of parameters.
It is certainly important to elucidate the generality of the result, e.g.,
 for other kinds of driving force and for other quantum-fluid systems,
 because then QT transitions may be characterized by the single
 coarse-grained equation of the DP class \cite{Hinrichsen-AP2000}.
We also find the unusual suppression
 of $\rho_\mathrm{ss}(V)$ near $V_\mathrm{c}$ (\figref{fig4}a), which seems
 to be related to the length-distribution of quantum vortices (\figref{fig2}b).

We hope that our results will trigger experimental studies
 of QT transitions along the DP-class scenario.
Experimentally, measurement of hysteresis may be a useful probe
 \cite{Takeuchi-PRE2008},
 which can circumvent the need for direct observation of quantum vortices.
We note that hysteretic behavior has indeed been reported
 for superfluid helium driven by oscillatory obstacles
 \cite{Hanninen.etal-PRB2007,Bradley.etal-PRB2014}.
Finally, we remark that the DP-class transition was also found, very recently,
 at the laminar-turbulent transition of classical sheared fluid
 \cite{Sano.Tamai-NP2016,Lemoult.etal-NP2016}.
Our results therefore add another link between QT and classical turbulence,
 unveiling nontrivial universal scaling laws behind,
 which arise in particular
 in the Kolmogorov-to-DP two-step relaxation reported here.


\section*{Methods}

\textbf{Numerical algorithm.}

We numerically integrated the GP equation \pref{eq:GP}
 in a cube of linear size $L=64$ (unless otherwise stipulated)
 with the periodic boundary condition.
Simulations were carried out by the fourth-order Runge-Kutta method
 with time step 0.01 and by the pseudospectral method
 using $M$ grid points for each dimension
 and cutoff wave number $k_\mathrm{cut} = \pi M/L$.
We chose $M=L$ except for quench simulations.
The random potential $V_{\vecr,t}$ was created as follows.
First, every $t_\mathrm{v}$ unit time, random numbers were generated
 from the uniform distribution in the range $[-V,V]$
 at every $\ell_\mathrm{v}$ unit length in 3-dimensional space.
These numbers were then smoothly connected by cubic spline curves
 with periodic boundary condition, so that the potential profile
 is continuous up to the second derivative.
The random potential was connected similarly in time,
 by assuming the vanishing first derivative at each reference time point,
 so that the random potential is continuous in time up to the first derivative.

To measure the vortex density, quantum vortices were counted
 plaquette by plaquette, i.e., the smallest closed loop passing
 four neighboring grid points.
Along each plaquette, we computed the phase difference
 $\Delta\theta = \sum_{i=0}^3 [\theta(\vecr_{i+1},t) - \theta(\vecr_i,t)]$,
 where $\vecr_0, \vecr_1, \vecr_2, \vecr_3, \vecr_4 = \vecr_0$ are
 the four grid points along the plaquette
 and $\theta(\vecr,t)$ is the phase of the wave function $\psi(\vecr,t)$.
If and only if $|\Delta\theta|>\pi$,
 we judged that there existed a vortex passing through this plaquette
($|\Delta\theta|$ should be a multiple of $2\pi$ but we used $|\Delta\theta|>\pi$ to admit numerical errors).
The vortex density $\rho$ was then given
 by the total number of such occupied plaquettes,
 divided by the system volume $L^3$.
Although a cell (box surrounded by six plaquettes)
 penetrated by a single vortex
 has two occupied plaquettes,
 vortices are not counted twice,
 because each plaquette is shared by two cells.

To compute the length of a vortex, 
 the entire vortex,
 i.e., a closed loop passing only through occupied plaquettes,
 should be constructed.
This can be easily done
 if each cell contains at most a single vortex, hence two occupied plaquettes.
If there are multiple vortices (or fragments of a vortex) in a single cell,
 one cannot determine pairs of plaquettes uniquely.
In our work, we chose the ``first-come-first-served'' basis;
 in other words, we simply connected two occupied plaquettes
 that we detected earlier.
Note that the fraction of such dense cells is very small in our simulations
 (see typical values of the vortex density $\rho$).

\textbf{Steady-state simulations.}

Our steady-state simulations started from a uniform initial state
 $\psi(\vecr,0)=1$, then the first 100 unit time was discarded as a transient.
We checked that different realizations reach the same steady state at $t=100$.
We then started to record quantities of interest.
Statistical data were obtained from several independent realizations,
 whose total time amounts to 3900 to 244000 excluding the discarded transients.
See Supplementary Video 1
 for typical evolution of vortices in such steady-state simulations.

\textbf{Measurement of local nucleation rate $h$.}

The nucleation rate $h$ was measured as follows.
For each realization,
 we started from the uniform initial state $\psi(\vecr,0)=1$
 with the given potential amplitude $V$,
 and measured the first nucleation time $t_\mathrm{n}$,
 at which the first vortex was generated.
This was repeated from 10000 to 300000 times for each $V$
 to obtain a histogram of $t_\mathrm{n}$.
The histogram was found to be exponential.
By fitting it to $\exp(-h(V)L^3 t_\mathrm{n})$,
 we evaluated the local nucleation rate
 $h(V)$ as a function of $V$.

\textbf{Quench simulations.}

The initial condition for the quench simulations was prepared as follows.
Similarly to the steady-state simulations,
 we first prepared a steady state in a fully developed turbulence regime,
 specifically at $V=140$.
Then we took $\psi(\vecr,t)$ at $t=500$ and used it
 as the initial condition for all quench simulations presented here.
Statistical data were obtained from at least $30$
 realizations for each $V$.
To resolve dense vortices in the initial condition and the following
 short-time regime of decay processes,
 here we used higher spatial resolution $M=3L=192$.
See Supplementary Video 2 for typical evolution of vortices
 in a quench simulation at $V=V_\mathrm{c}=10.5$.

\section*{Acknowledgements}

The authors thank M. Tsubota for useful discussions.
We acknowledge MEXT-supported Innovative Area ``Fluctuation \& Structure''
 which gave birth to this joint work,
 as well as its associated JSPS KAKENHI Grant Numbers
 JP25103004, JP25103007, JP26103519.
This work is also supported in part by
 JSPS KAKENHI Grant Numbers JP25707033 and JP16K21345,
 as well as the JSPS Core-to-Core Program
 ``Non-equilibrium dynamics of soft matter and information''.

\section*{Author contributions}

M.K. and K.A.T. designed the project.
M.T. and M.K. made simulation programs.
M.T. carried out simulations.
M.T. and K.A.T. analyzed the data.
All authors discussed the results.
M.T. and K.A.T. wrote the paper.

\section*{Additional information}

Correspondence should be addressed to K.A.T.
Requests for materials should be addressed to M.T.

\section*{Competing financial interests}

The authors declare no competing financial interests.



\bibliographystyle{naturemag}
\bibliography{DPref}

%

\end{document}